\documentclass[preprintnumbers,superscriptaddress,showpacs,pra]{revtex4}
\usepackage{amsmath}
\usepackage{epsfig}
\usepackage{graphicx}
\usepackage{color}

\input{tcilatex}
\begin{document}

\title{The centripetal force law and the equation of motion for a particle
on a curved hypersurface}
\author{L. D. Hu}
\affiliation{School for Theoretical Physics, School of Physics and Electronics, Hunan
University, Changsha 410082, China}
\author{D. K. Lian}
\affiliation{School for Theoretical Physics, School of Physics and Electronics, Hunan
University, Changsha 410082, China}
\author{Q. H. Liu}
\email{quanhuiliu@gmail.com}
\affiliation{School for Theoretical Physics, School of Physics and Electronics, Hunan
University, Changsha 410082, China}
\date{\today }

\begin{abstract}
It is pointed out that the current form of extrinsic equation of motion for
a particle constrained to remain on a hypersurface is in fact a
half-finished version for it is established without regard to the fact that
the particle can never depart from the geodesics on the surface. Once the
fact be taken into consideration, the equation takes that same form as that
for centripetal force law, provided that the symbols are re-interpreted so
that the law is applicable for higher dimensions. The controversial issue of
constructing operator forms of these equations is addressed, and our studies
show the quantization of constrained system based on the extrinsic equation
of motion is favorable.
\end{abstract}

\maketitle

\section{Introduction}

The study of the centripetal force law (CFL) can be traced back at least to
Newton in 1684 in his manuscript entitled "On the motion of bodies in an
orbit", \cite{newton} which was crucial step toward his law of universal
gravitation. In classical mechanics, it is the gravity that provides the
centripetal force responsible for astronomical orbits. However, in modern
physics, there is no gravity but the curved space-time, and the astronomical
orbits are nothing but geodesics in it. Though there is no gravity but
geodesics, is there any force law that bears resemblance to the CFL for an
orbit on a curved space?\ If yes, is it any beneficial to resolve the
problem associated with the quantization of the constrained system?

For the motion of a particle on a curved hypersurface is an exactly solvable
model to examine various problems such as higher-dimensional gravity, \cite%
{flat} dark energy/matter problem, \cite{liu13} and quantization of
constrained motions, \cite{homma,ikegami,weinberg} etc., \cite{cone} we are
familiar with both the geodesic equation from the intrinsically curved
surface and the equation of motion from the extrinsically Euclidean space, 
\cite{ikegami,weinberg} but no relationship in between has been seriously
explored. From the point of geometry, the geodesic equation can be
independent from the extrinsic world, but the extrinsic equation of motion
for a particle can never be independent from the geodesics because the
particle can not move unless follows one of them.

For a non-relativistic, mass $\mu $, particle that is constrained to remain
on a hypersurface described by a constraint $f(\mathbf{r})=0$ in $N$
dimensional Euclidean space $E^{N}$ spanned by $N$ \ mutually orthogonal
unit vectors $\mathbf{e}_{i}$ ($i=1,2,...,N$), where $f(\mathbf{r})$ is some
smooth function of position $\mathbf{r}=x^{i}\mathbf{e}_{i}$ and the
Einstein summation convention of sum over repeated indices is hereafter
assumed, there are two equations of motion for the particle. One is
well-known, given by the differential equation for a geodesic line $C$
determined by, 
\begin{equation}
\frac{d^{2}u^{\mu }(s)}{ds^{2}}+\Gamma _{\alpha \beta }^{\mu }\frac{%
du^{\alpha }(s)}{ds}\frac{du^{\beta }(s)}{ds}=0,  \label{1}
\end{equation}%
where $\{u^{\mu }\}$ ($\mu =1,2,...N-1$) are the intrinsic $N-1$ local
coordinates, and $s$ stands for the arc-length along $C$, and $\Gamma
_{\alpha \beta }^{\mu }$ are Christoffel symbols of the second kind. Another
is quite known, \cite{ikegami,weinberg} given by the differential of
velocity $\mathbf{v}\equiv d\mathbf{r}/dt$ with respect to time $t$,%
\begin{equation}
\frac{d}{dt}\mathbf{v=}-\mathbf{n}\left( \mathbf{\mathbf{v\cdot }\nabla 
\mathbf{n\cdot v}}\right) \mathbf{,}  \label{2}
\end{equation}%
where $\mathbf{n}\equiv \nabla f(\mathbf{r})/\left\vert \nabla f(\mathbf{r}%
)\right\vert $ is the local unit normal vector on the surface at point $%
\mathbf{r}=\left( x^{1}(\{u^{\mu }\}),x^{2}(\{u^{\mu }\}),x^{3}(\{u^{\mu
}\}),...,x^{N}(\{u^{\mu }\})\right) $, and $\nabla \equiv \mathbf{e}%
^{i}\partial /\partial x^{i}$ is the usual gradient operator. Though both
equations (\ref{1}) and (\ref{2}) share a salient feature that neither
contains the mass $\mu $, the possible difference between them is more
interesting and challenging. Some authors claim both to be \textit{identical 
}but without justification. \cite{ikegami} In order to obtain a proper form
of Eq. (\ref{2}) in quantum mechanics, it is usually assumed \ \cite%
{weinberg,liu16} that the equation (\ref{2}) has direct correspondence in
quantum mechanics once the velocity is rewritten in terms of momentum $%
\mathbf{v}=\mathbf{p/}\mu $, with possible ordering distributions of the
momentum $\mathbf{p}$ and position-dependent functions $\mathbf{\nabla 
\mathbf{n}}$ and $\mathbf{\mathbf{n}}$, in Heisenberg picture,%
\begin{equation}
\frac{d}{dt}\mathbf{p=}-\mathbf{n}\left( \frac{\mathbf{\mathbf{p}}\cdot 
\mathbf{\nabla \mathbf{n}}\cdot \mathbf{\mathbf{p}}}{\mu }\right) .
\label{2.1}
\end{equation}%
However, on one hand, Eq. (\ref{1}) is purely from intrinsic geometry, from
which we know that the Dirac\ quantization of constrained systems can not be
fulfilled throughout. \cite{liu14} On the other hand, Eq. (\ref{2.1})
contains mutually dependent components of the momentum $\mathbf{\mathbf{p}}$
because the motion lies on the tangential plane to the surface so that $%
\mathbf{\mathbf{n}}\cdot \mathbf{\mathbf{p}}=0$, thus in quantum mechanics,
we have inequivalent forms of (\ref{2.1}). \cite{liu16} In other words, Eq. (%
\ref{1}) under-describes the motion of the particle, which must be enlarged,
while the Eq. (\ref{2.1}) over-describes it, which must be used with some
constraints. Therefore, a proper form of equation of motion for the particle
in classical mechanics is worthy to be investigated. Whereas in quantum
mechanics the meaning of Eq. (\ref{2.1}) is under dispute,\ Ikegami \textit{%
et. al.} \cite{ikegami} from a larger sense concluded that from the
constraint equation $f(\mathbf{r})=0$ we could not build up a satisfactory
theory and we must start from another constraint equation $df(\mathbf{r}%
)/dt=0$,  but Weinberg \cite{weinberg} thought that Eq. (\ref{2.1}) should
be as true as it is in classical mechanics. Though no explicit use of the
Dirac formalism for a constrained system in present paper, our explorations
are in fact within it because we further develop the results (\ref{1})-(\ref%
{2.1}) given by it.

In the following section II, the generalized form of the CFL which
incorporates both Eqs. (\ref{2}) and (\ref{2.1}) is given. In section III,
the quantization problem of the constrained motion on the surface is
addressed. In final section IV, a brief conclusion and discussion is
presented.

\section{The generalized CFL unifies both intrinsic and extrinsic equation
of motion}

For our purpose, let us first recall the celebrated CFL $a=v^{2}/r$ for the
particle moves on a planar curve, especially on the $2D$ circle of radius $r$%
, and it can readily be rewritten in terms of the curvature $%
\kappa(\equiv1/r)$ and the Hamiltonian $H\equiv p^{2}/2\mu=\mu v^{2}/2$ for
the free motion without any external force imposed, 
\begin{equation}
\frac{d}{dt}\mathbf{p=}-2H\kappa\mathbf{n}.   \label{3}
\end{equation}
In fact, the Eq. (\ref{3}) holds true in general provided that $\kappa$
symbolizes the\ first curvature of the geodesic $C$ on the hypersurface and
Hamiltonian $H\ $applies to the free particle on the surface.

From the differential geometry for the hypersurface, at the point $\{u^{\mu
}\}$ on the surface $f(\mathbf{r})=0$, we can define the vectors of
tangential space $\mathbf{r}_{\alpha }(\equiv d\mathbf{r}/du^{\alpha }$) and
the unit normal vector $\mathbf{n}$, and these vectors $\left\{ \mathbf{r}%
_{\alpha },\mathbf{n}\right\} $ form a complete set of the coordinates in
the vicinity of the surface in the $E^{N}$, other than the fixed Cartesian
one $\left\{ \mathbf{e}_{i}\right\} $. The first and second fundamental
quantities are $g_{\alpha \beta }\equiv \mathbf{r}_{\alpha }\cdot \mathbf{r}%
_{\beta }$ and $b_{\alpha \beta }\equiv \mathbf{r}_{\alpha \beta }\cdot 
\mathbf{n}=-\mathbf{r}_{\alpha }\cdot \mathbf{n}_{\beta }$, respectively.
The equations of motion for $\mathbf{r}_{\alpha }$ and $\mathbf{n}$ are in
the $E^{N}$, \cite{dgs} 
\begin{subequations}
\begin{align}
\frac{\partial \mathbf{r}_{\alpha }}{\partial u^{\beta }}& =\Gamma _{\alpha
\beta }^{\mu }\mathbf{r}_{\mu }+b_{\alpha \beta }\mathbf{n},  \label{4} \\
\frac{\partial \mathbf{n}}{\partial u^{\alpha }}& =\mathbf{-}b_{\alpha
}^{\beta }\mathbf{r}_{\beta }.  \label{5}
\end{align}%
Furthermore, from the differential geometry for curves $\mathbf{r}(s)$ lying
on the surface, we can define, respectively, the unit tangential $\mathbf{%
\alpha }$ and its derivative with respect to $s$ in the following, 
\end{subequations}
\begin{subequations}
\begin{align}
\mathbf{\alpha }& \equiv \frac{d\mathbf{r}(s)}{ds}=\frac{\partial \mathbf{r}%
}{\partial u^{\alpha }}\frac{du^{\alpha }}{ds}\equiv \mathbf{r}_{\alpha }%
\frac{du^{\alpha }}{ds},  \label{6} \\
\frac{d^{2}\mathbf{r}}{ds^{2}}& =\frac{d\mathbf{\alpha }}{ds}=\frac{d}{ds}%
\left( \mathbf{r}_{\alpha }\frac{du^{\alpha }}{ds}\right) .  \label{7}
\end{align}

First, we limit the curve $\mathbf{r}(s)$ to be the geodesic line $C$. On
one hand, the first curvature $\kappa$ is defined by, \cite{dgc} 
\end{subequations}
\begin{equation}
\frac{d^{2}\mathbf{r}}{ds^{2}}=\kappa\mathbf{m,}   \label{8}
\end{equation}
where vector $\mathbf{m}$ is a unit normal vector of the curve, which by the
convention of the geometry is identical to $-\mathbf{n}$ (Recall that any
normal section of a surface is a geodesic \cite{dgs,dgc}). On the other
hand, the right-handed side of the Eq. (\ref{7}) becomes, from (\ref{4}) and
(\ref{1}), 
\begin{subequations}
\begin{align}
\frac{d}{ds}\left( \mathbf{r}_{\alpha}\frac{du^{\alpha}}{ds}\right) & =\frac{%
\partial\mathbf{r}_{\alpha}}{\partial u^{\beta}}\frac{du^{\alpha}}{ds}\frac{%
du^{\beta}}{ds}+\mathbf{r}_{\alpha}\frac{d^{2}u^{\alpha}(s)}{ds^{2}}
\label{9} \\
& =(\Gamma_{\alpha\beta}^{\mu}\mathbf{r}_{\mu}+b_{\alpha\beta}\mathbf{n})%
\frac{du^{\alpha}}{ds}\frac{du^{\beta}}{ds}-\Gamma_{\alpha\beta}^{\mu}\frac{%
du^{\alpha}(s)}{ds}\frac{du^{\beta}(s)}{ds}\mathbf{r}_{\mu}  \label{10} \\
& =b_{\alpha\beta}\frac{du^{\alpha}}{ds}\frac{du^{\beta}}{ds}\mathbf{n}. 
\label{11}
\end{align}
Substituting (\ref{8}) and (\ref{11}) into both sides of (\ref{7}), we get,
with noting $\mathbf{m=-n}$, the relation between curvature $\kappa$ and the
second fundamental quantities $b_{\alpha\beta}$, 
\end{subequations}
\begin{equation}
\kappa=-b_{\alpha\beta}\frac{du^{\alpha}}{ds}\frac{du^{\beta}}{ds}. 
\label{12}
\end{equation}
We see that the curvature $\kappa$ of the \textit{curve} is related to the
second fundamental quantities, the extrinsic geometric ones, of the \textit{%
surface}. In geometry, $-\kappa\mathbf{n}$ is a geometric invariant under
parameter transformation\ $\left\{ u^{\alpha}\right\} \rightarrow \left\{
u^{\prime\alpha}\right\} $.

Secondly, taking derivative of orthogonal relation $\mathbf{n}\cdot \mathbf{r%
}_{\gamma}=0$ with respect to any local coordinate $u^{\alpha}$ with noting $%
\mathbf{n}=\mathbf{n}(\mathbf{r}(u))$, we find, 
\begin{equation}
0=\frac{\partial}{\partial u^{\alpha}}\left( \mathbf{n\cdot\mathbf{r}}%
_{\gamma}\right) =\mathbf{r}_{\alpha}\cdot\nabla\mathbf{n}\cdot \mathbf{r}%
_{\gamma}+\mathbf{n\cdot}\frac{\partial}{\partial u^{\alpha}}\mathbf{r}%
_{\gamma}=\mathbf{r}_{\alpha}\cdot\nabla\mathbf{n}\cdot \mathbf{r}%
_{\gamma}+b_{\alpha\gamma},i.e.,b_{\alpha\gamma}=-\mathbf{r}%
_{\alpha}\cdot\nabla\mathbf{n}\cdot\mathbf{r}_{\gamma},   \label{13}
\end{equation}
where Eq. (\ref{4}) is used in simplifying $\mathbf{n\cdot}\partial \mathbf{r%
}_{\gamma}/\partial u^{\alpha}$. Substituting $b_{\alpha\beta }=-{\mathbf{r}%
_{\alpha}}\cdot\nabla{\mathbf{n}}\cdot\mathbf{r}_{\beta}$ into Eq. (\ref{12}%
), we have another form of the curvature $\kappa$,%
\begin{equation}
\kappa=-b_{\alpha\beta}\frac{du^{\alpha}}{ds}\frac{du^{\beta}}{ds}=\mathbf{r}%
_{\beta}\cdot\nabla\mathbf{n}\cdot\mathbf{r}_{\alpha}\frac{du^{\alpha}}{ds}%
\frac{du^{\beta}}{ds}=\frac{d\mathbf{r}}{ds}\cdot \nabla\mathbf{n}\cdot\frac{%
d\mathbf{r}}{ds}.   \label{14}
\end{equation}

Thirdly, since $\mathbf{\alpha}=d\mathbf{r/}ds$ is the unit tangential
vector along the curve $C$ and so is the ratio $\mathbf{p}/p=\mathbf{v}/v$,
the expression $\mathbf{p\cdot}\nabla\mathbf{n\cdot p/}\mu$ can be written
within the frameworks of the Hamiltonian mechanics and differential
geometry, which is given by,%
\begin{equation}
\frac{\mathbf{p}\cdot\nabla\mathbf{n}\cdot\mathbf{p}}{\mu}=\left( \frac{%
\mathbf{p}}{p}\cdot\nabla\mathbf{n}\cdot\frac{\mathbf{p}}{p}\right) \frac{%
p^{2}}{\mu}=\left( \frac{d\mathbf{r}}{ds}\cdot\nabla\mathbf{n}\cdot\frac{d%
\mathbf{r}}{ds}\right) \frac{p^{2}}{\mu}=2\kappa H.   \label{15}
\end{equation}
Substituting it into (\ref{2.1}), we in final reach the Eq. (\ref{3}). Thus,
the CFL (\ref{3}) really holds true universally.

For the particle is constrained on an arbitrary $N-1$ dimensional space
curve $f(\mathbf{r}(u(s)))=0$, the similar fashion gives the CFL (\ref{3})
as well. Under the coordinate transformation:\ $\left\{ u^{\alpha}\right\}
\rightarrow\left\{ u^{\prime\alpha}\right\} (\neq\left\{ u^{\alpha }\right\}
)$, two equations (\ref{1}) and (\ref{3}) are completely different for the
former transforms accordingly whereas the letter keeps invariant. In other
words, Eq. (\ref{3}) is geometric invariant whereas the Eq. (\ref{1}) is
not, though both are covariant.

\section{On the quantization problem of the constrained motion}

Why did we use the frameworks of Hamiltonian mechanics and differential
geometry to represent our result (\ref{3})? It not only formulates the
seemingly different physics laws into a compact and unified form, but also
sheds new light on the curvature-induced additional energy $V_{q}$ in
quantum mechanics. \cite{dacosta} In quantum mechanics, the equation of
motion (\ref{1}) takes following commutator version,%
\begin{equation}
\lbrack\mathbf{p},H]=i\hbar\left( \kappa\mathbf{n}H+H\kappa\mathbf{n}\right)
.   \label{16}
\end{equation}
Since $[\mathbf{x},H]=i\hbar\mathbf{p}/{\mu}$ gives the geometric momentum $%
\mathbf{p}$, \cite{liu13,liu11} this relation (\ref{16}) requires that
quantum Hamiltonian include an additional term $V_{q}$ such as $%
H=p^{2}/2\mu+V_{q}$, otherwise the equation (\ref{16}) would go violated.
For instance, for an $N-1$ dimensional sphere in $E^{N}$, the additional
energy is $V_{q}=(N-1)(N-3)\hbar^{2}/8mr^{2}$, \cite{liu13} which is exactly
the geometric potential predicted by the confining potential technique \cite%
{ikegami,dacosta} for the system under consideration. The additional energy
has been confirmed by experiments \cite{exp1,exp2} and may play some roles
in understanding of our present universe. \cite{liu13}

However, whether the Eq. (\ref{16}) holds true in general is dubious. This
is because Eq. (\ref{16}) contains the first curvature $\kappa $ of the
geodesic curve which depends on a single coordinate, which represents a
classical orbit, and its operator version is hard to be true except the
curvature is a constant. To reveal an even deeper difficulty, let us
consider a particular situation that the parametrization is so chosen that
the motion has unit speed (the so-called unit speed parametrization), the
coordinate parameterizing the geodesic can be taken as the time in classical
mechanics. But in quantum mechanics, the time remains a parameter without
operator counterpart while the coordinate could be quantized. So, Eq. (\ref%
{16}) can not take effect unless in classical limit.

Now let us turn to the Eq. (\ref{2.1}). As we stressed in the first section,
this equation over-describes the motion of the particle on the surface: \cite%
{liu16} The dependence of Eq. (\ref{2.1}) on components of position and
momentum is ambiguous, because we are free to choose not only independent
coordinates but also momenta for we have two constrained conditions $f(%
\mathbf{r})=0$ and $\mathbf{\mathbf{n}}\cdot \mathbf{\mathbf{p}}=0$. It may
not be a shortcoming, though. Instead, the over-description has a remarkable
advantage that includes the results predicted by the the confining potential
technique. \cite%
{liu13,ikegami,dacosta,liu16,liu14,liu11,liu13-2,liu13-3,liu15}    

Thus, we see that, even from the point of operator algebra, a complete
formulation of the quantization of the constrained motion is still an open
problem, \cite{homma,heller} our approach supports Weinberg whose point is
the Eq. (\ref{2.1}) holds true in quantum mechanics, \cite{weinberg} and
disfavors Ikegami \textit{et. al.} \cite{ikegami} whose point is that (\ref%
{2.1}) identical to Eq. (\ref{1}) and the satisfactory theory from
constraint equation $f(\mathbf{r})=0$ is not reachable with Dirac formalism
for quantizing a constrained system. 

\section{Conclusions and discussions}

The extrinsic equation of motion (\ref{2.1}) for the particle on a
hypersurface is only meaningful for the particle moves along the geodesic $C$%
, so in classical mechanics the current form (\ref{2.1}) of the equations in
literature is only a half-finished version. The final version of the
equation takes a compact form that turns out to be that for CFL once it is
generalized. Therefore, even there is no gravity but geodesics on a curved
space, there is a force law that bears striking resemblance to the CFL.
However, in quantum mechanics, the operator equation corresponding the CFL (%
\ref{3}) has a limited meaning, the equation corresponding (\ref{2.1}) opens
a wider door to establish a satisfactory quantum theory within the Dirac
formalism for quantizing a constrained system.

\begin{acknowledgments}
This work is financially supported by National Natural Science Foundation of
China under Grant No. 11175063.
\end{acknowledgments}

\end{document}